\documentstyle[amssymb,12pt]{amsart}
\newtheorem{thm}{Theorem}

\newtheorem{lem}[thm]{Lemma}
\newtheorem{prop}[thm]{Proposition}
\theoremstyle{remark}
\newtheorem{rem}{Remark}

\begin{document}

\newcommand{\thmref}[1]{Theorem~\ref{#1}}
\newcommand{\secref}[1]{\S\ref{#1}}
\newcommand{\lemref}[1]{Lemma~\ref{#1}}
\newcommand{\propref}[1]{Proposition~\ref{#1}}
\newcommand{\nc}{\newcommand}
\nc{\on}{\operatorname}
\nc{\ch}{\on{ch}}
\nc{\bw}{\bold{w}}
\nc{\onw}{\operatornamewithlimits}
\nc{\Z}{{\Bbb Z}}
\nc{\C}{{\Bbb C}}
\nc{\bbe}{\bold{1}}
\nc{\Pro}{{\Bbb P}}
\nc{\R}{{\Bbb R}}
\nc{\cond}{|\,}
\nc{\al}{\alpha}
\nc{\de}{\delta}
\nc{\til}{\tilde}
\nc{\bib}{\bibitem}
\nc{\la}{\lambda}
\nc{\th}{\theta}
\nc{\lar}{\longrightarrow}
\nc{\arr}{\rightarrow}
\nc{\fel}{\frac{1}{2}}
\nc{\maps}{\longmapsto}
\nc{\frwd}{\fracwithdelims()}
\nc{\ds}{\displaystyle}
\nc{\pone}{\Pro^1}

\title[Dilogarithm identities]{Dilogarithm identities, $q$-difference
equations \\ and the Virasoro algebra}

\author{Edward Frenkel}
\address{Department of Mathematics \\ Harvard University \\ Cambridge, MA
02138}
\thanks{Research of the first author was  supported by a Junior Fellowship
from the Harvard Society of Fellows and by NSF grant DMS-9205303}

\author{Andr\'as Szenes}
\address{Department of Mathematics \\ Massachusetts Institute of Technology
\\ Cambridge, MA 02139}
\thanks{E-mail addresses: frenkel@@math.harvard.edu, szenes@@math.mit.edu}

\date{December, 1992}
\maketitle

\section{Introduction}

In this paper we give a new proof of the identity
\begin{equation} \label{main}
 \sum_{j=1}^{n-1} L(\de_j) = \frac{\pi^2}{6} \frac{2n-2}{2n+1},
\end{equation}
where $L(z)$ is the Rogers dilogarithm function:
$$ L(z) = -\int_0^z \log(1-w)\,d\log w+\fel\log z \log(1-z),\quad 0\leq z\leq
1,$$
and $\de_j = \sin^2(\frac{\pi}{2n+1})/\sin^2(\pi \frac{j+1}{2n+1})$.

This equality was proved by Richmond and Szekeres from the asymptotic
analysis of the Gordon identities \cite{rsz} and also by Kirillov using
analytic methods \cite{kir}. There is a wide class of identities of this
type, which emerged in recent works on two-dimensional quantum field
theories and statistical mechanics. They appear in the calculation of the
critical behavior of integrable models, using the so-called Thermodynamic
Bethe Ansatz equation (\cite{kirresh,bazresh,kunnak,nahm,zam,klmer}). This
is an elegant, but mathematically non-rigorous method.

In this paper, we present a new approach to the identities, which relies on
the concepts of quantum field theory, and is rigorous at the same time. We
hope this will lead to a better understanding of the mathematical
structures behind general integrable two-dimensional quantum field
theories.

The idea goes back to Feigin \cite{feigin}: There is a certain increasing
filtration on each irreducible minimal representation of the Virasoro
algebra (or another conformal algebra) by finite-dimensional subspaces, whose
characters  are connected by a $q$-difference equation.  This
equation can be used to obtain an expression for the asymptotics of the
character of this representation in terms of values of the dilogarithm. On
the other hand, it is known that this asymptotics is determined by the
(effective) central charge $c_{\on{eff}}$ (cf. \cite{kacw}), and this gives
us an identity.

To illustrate that the values of the dilogarithm
should appear in the asymptotics of the solutions to $q$-difference
equations, consider the equation
$$ f(qx) = (1-aq)f(x), \quad 0<q<1.$$
Then
\begin{equation} \label{riemann}
\lim_{n\arr\infty} f(q^n) = f(1) \prod_{n=1}^\infty(1-aq^n).
\end{equation}
Let us calculate the asymptotics of this solution as $q\arr 1$.
We have
$$ \log q \log f(q^n) = \log q (\log f(1)+\sum_{n=1}^\infty \log(1-aq^n)).$$
This expression can be interpreted as a Riemann sum. Thus in the limit
$q\arr 1$ it gives
$$-\int_0^1 \log(1-ax)\,d\log x = L(a) - \fel \log a\log(1-a).$$

We will show that this approach can be carried out for the irreducible
representations of the $(2,2n+1)$ models of the Virasoro algebra. In this
case $c_{\on{eff}}=(2n-2)/(2n+1)$, and the corresponding identity is
\eqref{main}.

Let us remark that according to Goncharov \cite{goncharov}, the dilogarithm
identities of this type define elements of torsion in $K_3(\R)$.

\section{$(2,2n+1)$ minimal models of the Virasoro algebra}

The Virasoro algebra is the Lie algebra generated by elements $\{L_i, i\in\Z\}$
and $C$, with the relations
$$ [L_i,L_j] = (i-j) L_{i+j} + \frac{1}{12} (i^3-i) \de_{i,-j} C
\quad \text{and} \quad [L_i,C] = 0.$$

Fix an integer $n>1$. Let $\bbe_n$ be the one-dimensional representation of
the subalgebra generated by $\{C,L_i, i\geq -1\}$, on which the elements
$L_i$ act by 0, and $C$ acts by multiplication by $1-3(2n-1)^2/(2n+1)$.
Denote by $\til{V}_n$ the induced module of the Virasoro algebra. It is
$\Z$-graded with respect to the action of $L_0$, $\deg(L_{-i}) =i$. By
the Poincar\'e-Birkhoff-Witt theorem this module has a $\Z$-graded basis
consisting of the monomials $\{L_{-m_1}\dots L_{-m_p}v\cond m_1\geq
m_2\geq\dots \geq m_p>1\}$, where $v$ is the generator of $\bbe_n$. The
module $\til{V}_n$ has a unique vector $u$ of degree $2n+1$, such that
$L_i u=0$ for $i >0$. The quotient $V_n$ of the module $\til{V}_n$ by the
submodule generated by $u$ inherits the $\Z$-grading, and is irreducible
\cite{kac,fefu}. It is called the vacuum representation of the
$(2,2n+1)$ minimal model of the Virasoro algebra \cite{bpz}.

Let $C_n = \{(m_1,\dots,m_p)\cond m_1\geq\dots \geq m_p>1, m_i\geq
m_{i+n-1}+2\}$.

\begin{prop}[\cite{feno}] The image of the set
$$ \{L_{-m_1}\dots L_{-m_p}v\cond (m_1,\dots,m_p)\in C_n\}$$
under the homomorphism $\til{V}_n\arr V_n$ gives a linear basis of $V_n$.
\end{prop}

\begin{rem} While the argument in \cite{feno} relied on the Gordon
identities, this statement can be proved directly, using only the
representation theory of the Virasoro algebra (see \cite{prep}).
\end{rem}

For every integer $N>0$, introduce the subspaces $W^r_N,\;0\leq r<n$, of
the module $V_n$, which are linearly spanned by the vectors $$
\{L_{-m_1}\dots L_{-m_p}v\cond (m_1,\dots,m_p)\in C_n, m_1\leq N, m_{r+1}
\leq N-1\}.$$

For any $\Z$-graded vector space $V=\oplus_{m=0}^\infty V(m)$ with $\dim
V(m)<\infty$, let $\ch V = \sum_{m=0}^\infty \dim V(m) q^m$, which is
called the character of $V$.  Denote by $w_N^r$ the character of
$W^r_N$, and introduce $\bw_N = (w_N^0,\dots,w_N^{n-1})$ and
$\bw_0=(0,\dots,1)$.

\begin{lem} We have the following recursion relation:
$$  \bw_N = M_n(q^N) \bw_{N-1},$$
where
$$ M_n(x)= \begin{pmatrix}
0 & 0 & \hdots & 0 & 1 \\
0 & 0 & \hdots & x & 1 \\
\hdotsfor{5} \\
0 & x^{n-2} & \hdots & x & 1  \\
x^{n-1} & x^{n-2} & \hdots & x & 1
\end{pmatrix}.$$
\end{lem}

This gives us the following formula for the character of $V_n$:
\begin{equation} \label{char}
 \ch V_n = \bw_0^t \prod_{N=1}^\infty M_n(q^N)\bw_0 =
\bw_0^t \dots M_n(q^N) \dots M_n(q^2)M_n(q) \bw_0 .
\end{equation}
Thus $\ch V_n$ is equal to $\bw_0^t \lim_{n\arr\infty} {\bold f}(q^n)$, where
${\bold f}(x)$ is the solution to the $q$-difference equation ${\bold
f}(qx)=M_n(x){\bold f}(x)$ with the initial condition ${\bold f}(1)=\bw_0$.

The character of $V_n$ is a function in $q$ for $0<q<1$. We will study its
asymptotics as $q\arr 1$.

The following result was proved by Kac and Wakimoto, and follows from the
modular properties of $\ch V_n$.

\begin{prop}[\cite{kacw}] \label{kacform}
$$ -\lim_{q\arr1} \log q \log \ch V_n = \frac{\pi^2}{6}\frac{2n-2}{2n+1}.$$
\end{prop}

\begin{rem} The number $\frac{2n-2}{2n+1}$ is
called the effective central charge of the  $(2,2n+1)$ minimal
model.
\end{rem}

In the next section, we will use \eqref{char} to derive another
expression for this asymptotics.

\section{Asymptotics of the infinite product}

The following lemma relates the asymptotics of \eqref{char} to the
asymptotics of the infinite product of the highest eigenvalues of
$M_n(q^N)$.

\begin{lem} \label{dlem}
 Denote by $d_n(x)$ the eigenvalue of $M_n(x)$ of maximal
absolute value.

\begin{enumerate}
\item[(a)]  For any x, $0<x\leq 1$, the matrix $M_n(x)$ has $n$ different real
eigenvalues and $d_n(x)>0$.

\item[(b)] The asymptotic behavior of $\ch V_n$ is the same as that of
      the infinite product of the $d_n(q^N)$, more precisely
\end{enumerate}
\begin{equation} \label{predel}
 -\lim_{q\arr1} \log q \log \ch V_n = -\lim_{q\arr1}
\log q\log \prod_{N=1}^\infty d_n(q^N).
\end{equation}

\end{lem}

Similarly to the calculation of \eqref{riemann}, the right hand side of
\eqref{predel} can be written as the following integral:
\begin{equation} \label{integral}
\int_0^1 \log d_n(x)\, d\log x.
\end{equation}
The rest of this section is devoted to the calculation of this integral.

Denote by $Q_n(\la,x)$ the characteristic polynomial of $M_n(x)$.
Our computation is based on an explicit rational parametrization of
the curve $Q_n(\la,x)=0$.

\begin{rem} On this curve, $\la$ and $x$ define two algebraic functions.
Our integral \eqref{integral} is the integral of $\,\log \la\,d\log x\,$
along a path on the curve. We would like to stress that it is the
rationality of this curve, that allows us to express this integral in terms
of values of the dilogarithm function.
\end{rem}

Introduce the $n\times n$ matrix
$$ L_n(x) = \begin{pmatrix}
1+x & -1 & 0 & \hdots & 0 & 0 & 0 \\
-x & 1+x & -1 & \hdots & 0 & 0 & 0 \\
0 & -x & 1+x & \hdots & 0 & 0 & 0 \\
\hdotsfor{7} \\
0 & 0 & 0 & \hdots & 1+x & -1 & 0 \\
0 & 0 & 0 & \hdots & -x & 1+x & -1 \\
0 & 0 & 0 & \hdots & 0 & -x & x
\end{pmatrix} ,$$
and observe that $M_n(x)^{-2} = x^{-n}L_n(x).$
Then the roots of the characteristic polynomial
$P_n(\mu,x)=\det(L_n(x) - \mu)$ are connected to the roots of $Q_n(\la,x)$
via the formula $\la^2 = x^n/\mu$. On the other hand, we have the following
recursion for $P_n(x)$:
$$  P_n(\mu,x) = (1+x-\mu) P_{n-1}(\mu,x) - x P_{n-2}(\mu,x),$$
with initial conditions $P_0(\mu,x) = 1$ and $P_1(\mu,x) = x-\mu$.

Introduce $t=x/(x-\mu+1)^2$ and $u=1/(x-\mu+1)$. In these new variables
our recursion can be written as
$$ u^{-1} \frac{P_{n-1}}{P_n} =
\frac{1}{1-t\left(u^{-1}\frac{P_{n-2}}{P_{n-1}}\right)}.$$
This can be solved by the  continuous fraction
$$ u^{-1} \frac{P_{n-1}}{P_n} =
\cfrac{1}{1-
\cfrac{t}{1-
\cfrac{t}{\dots
\cfrac{\dots}{1-
\cfrac{t}{1-u}}}}}
. $$
As a consequence the equation $P_n=0$ is equivalent setting the denominator
of this continuous fraction to 0. This leads to the following parametrization
for $u$:
$$ u= A_n(t),\quad \text{where} \quad A_n(t) =
1-\cfrac{t}{1-
\cfrac{t}{\dots
\cfrac{\dots}{1-
\cfrac{t}{1-t}}}}
.$$

Naturally, $A_n(t)$ satisfies the recursion $A_n(t) = 1-t/A_{n-1}(t)$ with
the initial condition $A_1(t) =1$. A quick calculation shows that as
functions of the parameter $t$ our original variables $\la$ and $x$ take
the form $$ x_n(t) = \frac{t}{A_n^2(t)}\qquad \text{and} \qquad
\la_n(t) = \frac{A_1(t)A_2(t)\dots A_{n-1}(t)}{A_n^{n-1}(t)}.$$

\begin{lem}
For any $\al$, $0<\al\leq1$, there are $n$ real positive solutions
$t_1(\al)<\dots < t_n(\al)$ to the equation $x_n(t) = \al$. The numbers
$\la_n(t_1(\al)), \dots \la_n(t_n(\al))$ are the eigenvalues of the
matrix $M_n(\al)$, and
$$ d_n(\al) = \la_n(t_1(\al)).$$
\end{lem}

The lemma implies that our integral \eqref{integral} can be written as
\begin{equation} \label{newint}
 \int_{t=0}^{t=t_1(1)} \log \la_n(t)\, d\log x_n(t).
\end{equation}

Introduce the rational functions $f_i(t) = 1- A_{i+1}(t)/A_i(t).$
The following key formula can be proved by induction:
\begin{equation} \label{key}
f_i(t) = 
\frac{t^i}{(1-f_1(t))^{2(i-1)}\dotsm(1-f_{i-1}(t))^2}.
\end{equation}

In terms of the these functions our variables can be expressed as
follows:
$$ x_n = \frac{t}{(1-f_1)^2(1-f_2)^2\dots(1-f_{n-1})^2}
$$ 
$$\la_n = \frac{1}{(1-f_1)(1-f_2)^2\dots(1-f_{n-1})^{n-1}}  $$

Now we can calculate the integral \eqref{newint}.
We begin with a more general integral
$$  \int_{0}^{\gamma} \log \la_n(t)\, d\log x_n(t),$$
where $\gamma>0$ is such that $\la_n(t)$ and $x_n(t)$ have no poles on the
interval $[0,\gamma]$. Substituting the formulas above, we obtain
$$ -\int_0^\gamma \sum_{j=1}^{n-1} j\log(1-f_j(t))\, d\left(\log t
-\sum_{i=1}^{n-1} 2\log(1-f_i(t))\right)=$$
\begin{eqnarray*}
  \ds -\int_0^\gamma \sum_{j=1}^{n-1}  \log(1-f_j(t))\, d \log\left(
\frac{t^j}{\prod_{i=1}^{j-1}(1-f_i(t))^{2(j-i)}}\right)  + & \\
  \ds \int_0^\gamma \sum_{i,j=1}^{n-1} 2\min(i,j)   \log(1-f_j(t))\,
 d\log(1-f_i(t)).&
\end{eqnarray*}
Applying  formula \eqref{key} to the first summand, and partial
integration to the second, our integral takes the form
$$
 -\int_0^\gamma \sum_{j=1}^{n-1}   \log(1-f_j(t))\, d \log f_j(t) +
  \sum_{i,j=1}^{n-1} \min(i,j) \log(1-f_j(\gamma)) \log(1-f_i(\gamma)).
$$
Now by the definition of the Rogers dilogarithm, this can be written as
\begin{eqnarray} \label{final}
 \ds \sum_{j=1}^{n-1} L(f_j(\gamma)) - \fel \sum_{j=1}^{n-1}
\log f_j(\gamma)\log(1-f_j(\gamma)) + & \\  \ds \sum_{i,j}^{n-1}
 \min(i,j) \log(1-f_j(\gamma))\log  (1-f_i(\gamma)).& \nonumber
\end{eqnarray}
To complete our proof, we need to describe some properties of the numbers
$\de_i$ of \eqref{main}.

\begin{lem} Fix an integer $n>1$. The numbers $\de_i$, $1\leq i\leq n-1$,
satisfy the following properties:
\begin{enumerate}
\item[(a)] $f_i(\de_1) = \de_i$,
\item[(b)] $\de_i = \prod_{i=1}^{n-1} (1-\de_j)^{2\min(i,j)}$,
\item[(c)] $ \de_1 = A_n(\de_1)^2$, thus $x_n(\de_1)=1$. Moreover,
$t_1(1)=\de_1$.
\end{enumerate}
\end{lem}

Combining the calculation above with property (c), we see that the integral
\eqref{newint} equals to the expression
\eqref{final}, with $\gamma=\de_1$. Using property (a) this is simply
$$ \sum_{j=1}^{n-1} L(\de_j) -\fel \sum_{j=1}^{n-1} \log\de_j \log(1-\de_j)
+ \sum_{i,j}^{n-1} \min(i,j) \log(1-\de_i)\log(1-\de_j).$$

Finally, according to property (b), the last two terms cancel. Hence we obtain
the following

\begin{prop}
$$\int_0^1 \log d_n(x)\, d\log x = \sum_{j=1}^{n-1} L(\de_j).$$
\end{prop}

This result, together with \lemref{dlem} and \propref{kacform}, proves
identity \eqref{main}.

\begin{rem} We are certain that this method is applicable to other models
of conformal field theory.
\end{rem}

\noindent{\bf Acknowledgements}. We are grateful to B. Feigin for
sharing his ideas with us. We would like to thank A. Goncharov, V.~Kac,
A.~Levin and T.~Nakanishi for valuable discussions.

\bibliographystyle{amsplain}

\end{document}